\documentclass[aps, prd]{revtex4-2}
\usepackage{xcolor}
\usepackage{float}
\usepackage{amsmath,amsfonts,amssymb,epsfig,graphicx}
\usepackage{amsmath,mathrsfs}
\usepackage{bbold}
\usepackage{bm}
\usepackage{amsfonts}
\usepackage{amssymb}
\usepackage{slashed}
\usepackage{hyperref}
\usepackage[normalem]{ulem}
\usepackage{lineno}
\usepackage{comment}
\usepackage{afterpage}
\usepackage{titlesec}
\usepackage{booktabs}
\usepackage{multirow}
\usepackage{subcaption} 
\usepackage{setspace}

\titleformat{\section}{\normalfont\Large\bfseries}{\thesection}{1em}{}
\newcommand{\ket}[1]{\left|#1\right\rangle}
\newcommand{\bra}[1]{\left\langle#1\right|}

\newcommand{\ttbar}{t\bar{t}}

\newcommand{\mc}{\mathcal}

\newcommand{\II}{\mathbb{1}}

\def\Tr{\mbox{Tr}\,}
\newcommand{\pythia}{{\large\texttt{P}}{\small \texttt{YTHIA}}~}
\newcommand{\delphes}{\texttt{Delphes}~}
\newcommand{\madgraph}{\texttt{Madgraph}~}
\newcommand{\cc}{C_{2,1,2,-1}}
\newcommand{\CC}{C_{2,2,2,-2}}

\begin{document}
\title{Testing Bell inequalities and probing quantum entanglement at a muon collider}

\author{Alim \surname{Ruzi}}
\email[]{alim.ruzi@pku.edu.cn}
\affiliation{State Key Laboratory of Nuclear Physics and Technology, School of Physics, Peking University, Beijing, 100871, China}

\author{Youpeng \surname{Wu}}
\email[]{phsjcq@pku.edu.cn}
\affiliation{State Key Laboratory of Nuclear Physics and Technology, School of Physics, Peking University, Beijing, 100871, China}

\author{Ran \surname{Ding}}
\affiliation{State Key Laboratory of Nuclear Physics and Technology, School of Physics, Peking University, Beijing, 100871, China}

\author{Sitian \surname{Qian}}
\affiliation{State Key Laboratory of Nuclear Physics and Technology, School of Physics, Peking University, Beijing, 100871, China}

\author{Andrew Micheal \surname{Levin}}
\email[]{andrew.michael.levin@cern.ch }
\affiliation{State Key Laboratory of Nuclear Physics and Technology, School of Physics, Peking University, Beijing, 100871, China}

\author{Qiang \surname{Li}}
\email[]{qliphy0@pku.edu.cn}
\affiliation{State Key Laboratory of Nuclear Physics and Technology, School of Physics, Peking University, Beijing, 100871, China}

\begin{abstract}
A muon collider represents a promising candidate for the next generation of particle physics experiments after the expected end of LHC operations in the early 2040s. Rare or hard-to-detect processes at the LHC, such as the production of multiple gauge bosons, become accessible at a TeV muon collider. We present here the prospects of detecting quantum entanglement and the violation of Bell inequalities in $H \to ZZ \to 4\ell$ events at a potential future muon collider. We show that the spin density matrix of the Z boson pairs can be reconstructed using the kinematics of the charged leptons from the Z boson decays. Once the density matrix is determined, it is straightforward to obtain the expectation values of various Bell operators and test the quantum entanglement between the Z boson pair. Through a detailed study based on Monte-Carlo simulation, we show that the generalized CGLMP inequality can be maximally violated, and testing Bell inequalities could be established with high significance.
\end{abstract}

\maketitle

\section{Introduction} 
Quantum entanglement~\cite{Horodecki:2009zz} is one of the most distinctive and counter-intuitive features of quantum mechanics. Particles that have interacted in the past remain in an entangled state even when spatially separated. Particles in an entangled state exhibit correlations in their physical properties, such as spin polarization. A correlated quantum system may even lead to the violation of the Bell inequalities~\cite{Bell:1964kc}, which implies that the correlations cannot be explained by local hidden variables. 
Quantum entanglement can be readily observed in two-qubit system, such as a pair of spin-1/2 particles or a pair of photons. Experimental searches for quantum entanglement and violation of Bell inequalities have been successfully performed in two-outcome measurements with correlated photon pairs~\cite{Freedman:1972zza,Clauser:1978ng,Aspect:1982fx}. Other proposals have been made to test Bell inequalities in $e^+e^-$ collision events~\cite{Tornqvist:1980af}, in  charmonium decays~\cite{Baranov:2008zzb,Fabbrichesi:2024rec}, and more recently in $\ttbar$ events at hadron colliders ~\cite{Afik:2020onf, Severi:2021cnj,Larkoski:2022lmv, Aguilar-Saavedra:2022uye, Afik:2022dgh,Afik:2022kwm, Han:2023fci, Dong:2023xiw}. 

The study of the quantum entanglement and testing Bell inequality violation can be another new subject for the high-energy physics community. While detectors at high-energy colliders are not specifically designed to probe quantum entanglement, they have demonstrated surprising effectiveness in this task. This opens up exciting opportunities for novel measurements in quantum information science, as well as potential discoveries that could extend beyond the Standard Model. Recently, spin polarization of pairs of top and anti-top quarks has been measured and quantum entanglement is observed~\cite{CMS:2019nrx,CMS:2024hgo,ATLAS:2023fsd}. It has also been shown that Bell inequality is violated in the decays of B mesons at LHCb and Belle II experiment~\cite{Fabbrichesi:2023idl}. Interestingly, several new proposals made to test violation of Bell inequality, such as CGLMP (Collins-Gisin-Linden-Massar-Popescu) inequality~\cite{Collins:2002sun}, an optimized one for two-qutrit system. These proposals leverage gauge boson production from Higgs boson decay~\cite{Maina:2020rgd,Aguilar-Saavedra:2022mpg,Aguilar-Saavedra:2022wam,Aoude:2023hxv,Barr:2021zcp,Bi:2023uop,Fabbri:2023ncz,Marzola:2023oyv}, heavy lepton pair production at a lepton collider and other future collider~\cite{Ehataht:2023zzt,Ma:2023yvd,Gray:2021jij}, and also from vector boson scattering process~\cite{Morales:2023gow}.

Massive gauge bosons may provide an efficient way to explore quantum entanglement and test the CGLMP inequality because they can act as their own polarimeters (in case of W boson) and exhibit higher dimensionality in their entangled system. The corresponding spin polarizations  can be reconstructed from the angular distribution of the final leptons or jets. It has been argued that the observation of quantum entanglement and testing Bell inequality could be achieved in di-boson production in proton-proton collision as well as a lepton collider with relatively better significance by determining the polarization density matrix of the entangled two-qutrit system~\cite{Fabbrichesi:2023cev}. 

In this paper, we investigate quantum entanglement and violation of Bell inequality, the CGLMP inequality, at a Muon Collider through detailed Monte Carlo (MC) simulation. The Muon Collider recently received revived interest due to its potential to achieve Center of Mass (CM) energies to 1-10 TeV range or beyond. Thus a Muon Collider allows for the observation of processes that are challenging to detect at the current CM energy of the LHC. These processes may include double Higgs boson production through Vector Boson Fusion (VBF), top quark pair production in association with scalar and gauge boson, diboson production etc~\cite{MuonCollider:2022nsa, MuonCollider:2022xlm,Aime:2022flm,Delahaye:2019omf}.
  
We focus on the process of $Z$ boson pair productions through decays of Higgs bosons produced via VBF process at a TeV scale muon collider, $\mu^+\mu^-\to \nu_{\mu}\bar{\nu}_{\mu} H$, $H \to Z Z$, at three different CM Energies: $\sqrt{s} = 1\,\rm{TeV},\, 3\,\rm {TeV},\, 10\,\rm {TeV}$. We choose this process because of the cleanest signal and the insignificant background. We simulate one million signal events through \madgraph\cite{Alwall:2014hca} and use \delphes\cite{deFavereau:2013fsa} program for detector simulations. The angular distribution of the final leptons coming from Z boson decay can be obtained precisely. Finally, probing quantum entanglement and Bell inequality violation require the knowledge of the spin correlations between two Z bosons, which can be fully reconstructed from angular distribution of the final leptons in the Z boson rest frame.

\section{Theoretical Framework}
\subsection{Density matrix and observable quantity}

The density matrix for the joint $ZZ$ system, $\rho$ acts on the 9 dimensional Hilbert space defined by the three spin states of each Z boson.
Defining the z-axis as the direction of Z boson's 3-momentum, we can choose the eigenstates of the momentum operator \(J_z\) as the basis vectors of the spin space. The $Z$ boson pairs are produced from Higgs decay, and the spin component is conserved in the momentum direction in the CM frame. Therefore, the ZZ state can only lie in one of 3 joint states, i.e., \({\ket{l_1l_2}\in\{\ket{-+}, \ket{00}, \ket{+-}}\}\), where \(l_1\) and \(l_2\) are the spin states of two Z bosons. By definition, the density matrix of a two-qutrit ZZ system can be written as a tensor production
\begin{align}
    \rho=\sum p_{l_1l_2}\ket{l_1l_2}\bra{l_1l_2},
\end{align} 
where $p_{l_1l_2}\geq 0$ and $\quad\sum p_{l_1l_2}=1$.
  \begin{figure}[htb]     \centering
     \includegraphics[width=0.45\textwidth]{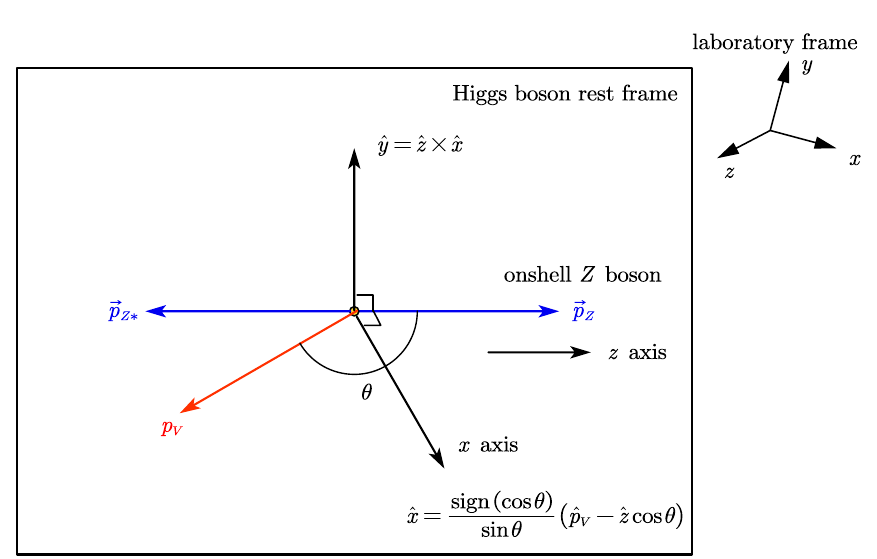}
     \includegraphics[width=0.45\textwidth]{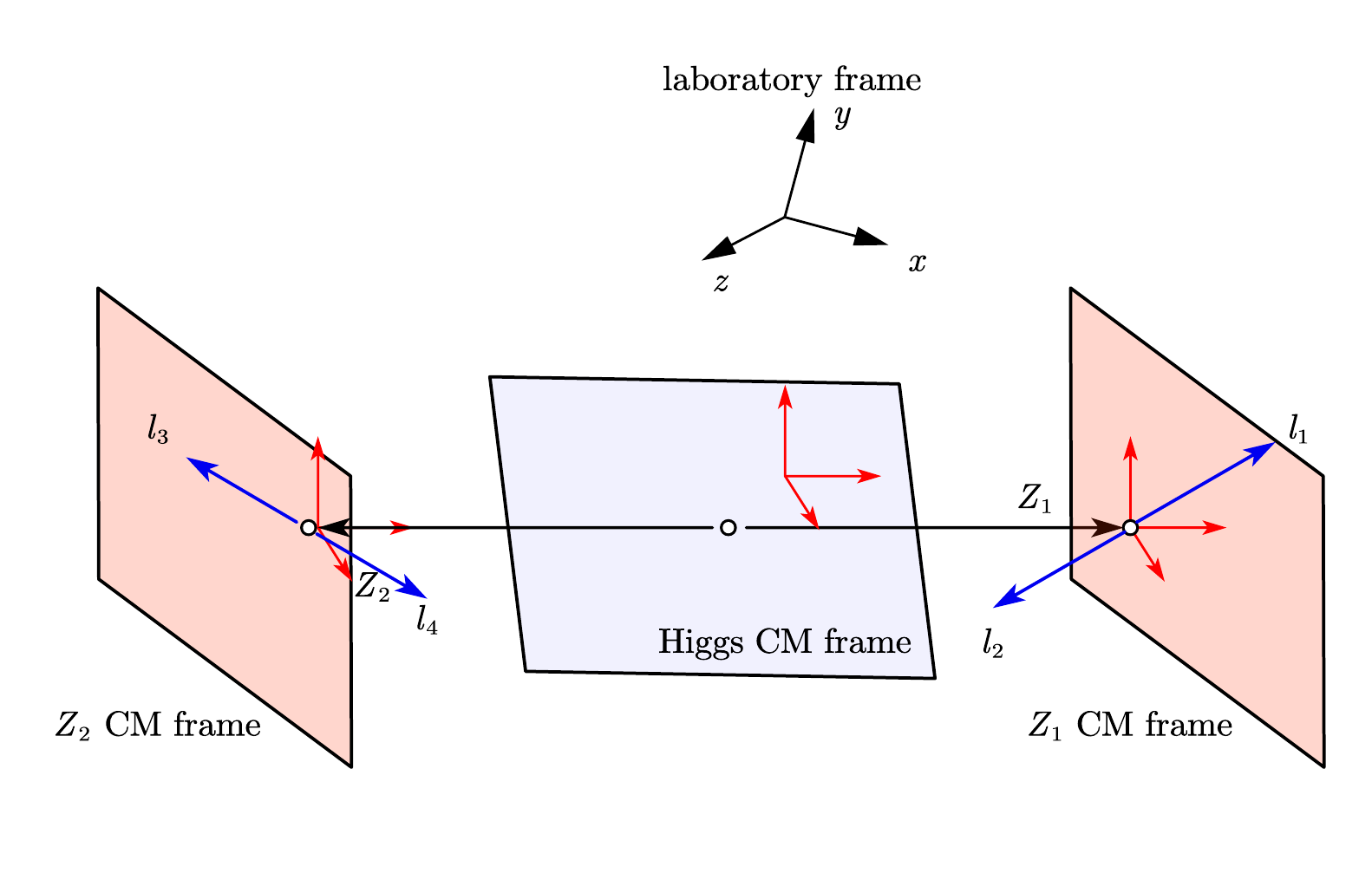}
     \caption{Definition of Reference System. In Higgs CM frame, we define on-shell bosons as z-axis so that we create coordinate. In the lepton CM frame, we will still use this coordinate system for the direction angles of the leptons. }
     \label{fig:frame}
 \end{figure}
 
Without the loss of generality, we choose to work in the CM frame, where the unit vectors and corresponding reference system is shown in Fig1.~\ref{fig:frame}.  The spin eigenstates of each Z boson is defined in their respective rest frame. Therefore, we need to boost $Z$ boson into the CM frame of the Higgs boson. The reference system in which the angular distribution of the decay leptons are determined is similar to the one applied in tp pair production~\cite{Bernreuther:2015yna} and massive gauge boson production in Higgs decay~\cite{Fabbrichesi:2023cev, Aguilar-Saavedra:2022wam}.
Firs of all, a unit vector $\hat{z}$ is defined in the three-momentum direction of the on-shell $Z$ boson in the Higgs boson rest frame. Secondly, the $\hat{x}$ vector on the production plane is defined as $\hat{x} = {\rm sign}(\cos\theta)(\hat{p}_V -\hat{z} \cos\theta )/\sin\theta$, with $\hat{p_V} = (0, 0, 1)$ the direction of the incident muon beams in the laboratory frame.
Finally, the  unit vector $\hat{y}$ is defined as $\hat{y} = \hat{z} \times \hat{x}$, perpendicular to the production plane.

The polarization vector of the spin-1 particle can be expressed as a linear combination of the three reference vectors as
\begin{align}
    \varepsilon^\mu(p,\lambda)=\frac{1}{\sqrt{2}}|\lambda|(\lambda n_1^\mu+\mathrm{i}n_2^\mu)+(1-|\lambda|)n_3^\mu.
    \label{eq:polarization}
\end{align} 
where the boosted vectors $n_{i}^{\mu}$ for each spin-1 particle are defined as
\begin{align}
    n_1^\mu(1)=n_1^\mu(2)=(0,\hat{y}), ~~~
    n_2^\mu(1)=n_2^\mu(2)=(0,\hat{x}), ~~~
    n_3^\mu(1)=\gamma(\beta,\hat{z}),~~~
    n_3^\mu(2)=\gamma(-\beta,\hat{z}),
\end{align}
with \(-\beta, \gamma\) boost parameters, and the indices \(1,2\) represent the two Z bosons. 
The probability amplitude of the \(ZZ\) productin with helicities $\lambda_1$ and $\lambda_2$ is given by 
\begin{align}
    \mathcal{M} \left( \lambda _1,\lambda _2 \right) =\mathcal{M} _{\mu \nu}\varepsilon ^{\mu}\left( p_1,\lambda _1 \right) \varepsilon^{\nu}\left( p_2,\lambda _2 \right),
    \label{eq:amplitude} 
\end{align}
The, the polarization density matrix of the $ZZ$ joint system can be written using the helicity basis as
\begin{align}    
\rho=\ket{\Phi}\bra{\Phi} &=\frac{1}{|\bar{\mathcal{M}}|^2}\mathcal{M}(\lambda_1,\lambda_2)\mathcal{M}(\lambda_1',\lambda_2')^*\\
& = \frac{1}{|\bar{\mathcal{M}}|^2} \left( \mathcal{M} _{\mu \nu}\varepsilon ^{\mu}\left( p_1,\lambda _1 \right) \varepsilon ^{\nu}\left( p_2,\lambda _2 \right) \right) \left( \mathcal{M} _{\mu' \nu'}\varepsilon ^{\mu'}\left( p_1,\lambda _1' \right) \varepsilon ^{\nu'}\left( p_2,\lambda _2' \right) \right) ^{\dag}.
\end{align}
where \(\bar{\mathcal{M}}\) is the spin-averaged unpolarized scattering amplitude. Using the polarization vector, we can construct the covariant helicity projection operator \(\mathscr{P}_{\lambda\lambda'}^{\mu\nu}\)as
\begin{align}
    \mathscr{P} _{\lambda \lambda '}^{\mu \nu}\left( p \right) &=\varepsilon ^{\mu}\left( p,\lambda \right) \varepsilon ^{\nu}\left( p,\lambda' \right),
\end{align}
and finally the density matrix for the joint system reads as 
\begin{align}
    \rho \left( \lambda _1,\lambda _{1}^{\prime},\lambda _2,\lambda _{2}^{\prime} \right) =\frac{\mathcal{M} _{\mu \nu}\mathcal{M} _{\mu \prime\nu \prime}^{\dagger}\mathscr{P} _{\lambda _1\lambda _{1}^{\prime}}^{\mu \mu \prime}\left( p_1 \right) \mathscr{P} _{\lambda _1\lambda _{1}^{\prime}}^{\mu \mu \prime}\left( p_2 \right)}{|\bar{\mathcal{M}}|^2}.
\end{align}
 
To obtain the density matrix from simulated or actual data, a proper form of parametrization is needed. The form of the density matrix describing the polarization state of the two-qutrit system formed by two spin-1 bosons can generally be parameterized using $3\times 3$ matrices composed of either Gell-Mann matrices ~\cite{Fabbrichesi:2023cev,Barr:2024djo} or spin-1 component operators. Using Gell-Mann matrices to represent the density matrix is one of the possible parameterization. There is another simple yet effective way to parameterize the density matrix composed of linear combination of irreducible tensor operators~\cite{Aguilar-Saavedra:2015yza, Aguilar-Saavedra:2017zkn, Rahaman:2021fcz}
\begin{equation}
    \rho = \frac19\left[
    \II_3\otimes\II_3 + A_{LM}^1 T_{M}^L\otimes\II_3 + A^2_{LM}\II_3\otimes T^L_{M} + C_{L_1M_1L_2M_2}T^{L_1}_{M_1}\otimes T_{M_2}^{L_2}
    \right],
    \label{eq:rho2}
\end{equation}
where $T^L_M$ are the irreducible tensor operators complying with a trace rule such that ${\rm Tr}\left[T_{M}^L (T^L_M)^\dagger\right]$ = 3. It should be noted that a sum over the indices $L = 1, 2$ and $-L \leq M \leq L $. These tensor operators are usually formed from linear combination of the spin  operators for spin-1 particles. One may refer to paper~\cite{Aguilar-Saavedra:2022wam} for the exact form of these operators. The coefficients $A^i_{LM}$ and $C_{L_1M_1L_2M_2}$ are   representing both  the polarizations and spin correlations of pair of Z bosons.  These coefficients are extracted from simulated or actual data, which is very challenging because both the CM frame of the colliding beam and the rest frame of the decaying particles should be established correctly so that the angular distributions of the final lepton can be determined precisely. Consequently, there will be multiple Lorentz boost or transformation between coordinate systems. 
In order to extract the coefficients of the density matrix, a relation between these coefficients and a direct observable in the experiment is needed. This has been done through a quantum state tomography for weak decays of massive bosons~\cite{Ashby-Pickering:2022umy}. The high energy collider experiments directly provide the general observable quantity, cross section, which can be expressed in terms of decaying matrix density of the massive bosons. For example, the differential cross section for $ZZ\to \ell_1^+\ell_1^-\ell_2^+\ell_2^-$ can be written~\cite{Rahaman:2021fcz}
\begin{equation}
    \frac{1}{\sigma} \frac{d\sigma}{d\Omega_+ \, d\Omega_-} = \left( \frac{3}{4\pi} \right)^2 \mathrm{Tr} \left[ \rho_{V_1 V_2} \left( \Gamma_1 \otimes \Gamma_2 \right) \right],
    \label{eq:xsection}
\end{equation}
where the solid angles $d\Omega^{\pm} \sin\theta^{\pm}d\theta^{\pm}d\phi^{\pm}$ are given in terms of the spherical coordinates for the momenta of the final charged leptons with respect to the rest frame of the decaying Z boson. $\rho_{V_1V_2}$ is the density matrix for the entangled $ZZ$ state explained previously. The decaying density matrix of a Z boson into charged leptons is given as~\cite{Rahaman:2021fcz,Aguilar-Saavedra:2022wam,Aguilar-Saavedra:2022mpg}
\begin{equation}
\Gamma(\theta, \phi) = \frac{1}{4} 
\begin{pmatrix}
1 + \cos^2 \theta - 2 \eta_\ell \cos \theta & \frac{1}{\sqrt{2}} (\sin 2\theta - 2\eta_\ell \sin \theta)e^{i\phi} & (1 - \cos^2 \theta)e^{i2\phi} \\
\frac{1}{\sqrt{2}} (\sin 2\theta - 2\eta_\ell \sin \theta)e^{-i\phi} & 2 \sin^2 \theta & -\frac{1}{\sqrt{2}} (\sin 2\theta + 2\eta_\ell \sin \theta)e^{i\phi} \\
(1 - \cos^2 \theta)e^{-i2\phi} & -\frac{1}{\sqrt{2}} (\sin 2\theta + 2\eta_\ell \sin \theta)e^{-i\phi} & 1 + \cos^2 \theta - 2 \eta_\ell \cos \theta
\end{pmatrix},
\end{equation}
where the spherical coordinates $\theta$ and $\phi$ are the angles of the three momentum of the negative  charged lepton in the $Z$ boson rest frame. The trace can be simplified further using the normalization property of the irreducible tensors and making use of spherical harmonic functions $Y_L^M(\theta, \phi)$ so that Eq.~\ref{eq:xsection} can be written in the following form:
\begin{align}
    \frac{1}{\sigma}\frac{d\sigma}{d\Omega_1d\Omega_2} =&\frac{1}{(4\pi)^2} [ 1+A_{LM}^1Y_L^M(\theta_1,\phi_1)+A_{LM}^2B_LY_L^M(\theta_2,\phi_2)\\
    & +C_{L_1M_1L_2M_2}B_{L_1}B_{L_2}Y_{L_1}^{M_1}(\theta_1,\phi_1)Y_{L_2}^{M_2}(\theta_2,\phi_2)
    ] ,
\end{align}
with $B_1 = -\sqrt{2\pi}\eta_{\ell}$, and $B_2 = \sqrt{2\pi/5}$.

Now the coefficients in Eq.~\ref{eq:rho2} can be obtained by integrating the above differential cross section over the solid angle of the leptons and using orthogonal property of the spherical harmonics:
\begin{align}
    \int \frac{1}{\sigma}\frac{d\sigma}{d\Omega_1d\Omega_2}Y_L^M(\Omega_j)d\Omega_j&=\frac{B_L}{4\pi}A_{LM}^j, \qquad j=1,2;\\
    \int \frac{1}{\sigma}\frac{d\sigma}{d\Omega_1d\Omega_2}Y_{L_1}^{M_1}(\Omega_1)Y_{L_2}^{M_2}(\Omega_1)d\Omega_1d\Omega_2 &= \frac{B_{L_1}B_{L_2}}{4\pi}C_{L_1M_1L_2M_2}.
    \label{eq:ACLM}
\end{align}

It is worth noting that the $ZZ$ system is in the singlet state because the third component of the spin along the boson momentum direction is conserved. This imposes strong constraints on the form of the density matrix, including only nine non-zero elements with the relation
\begin{equation}
    C_{2,2,2,-2} = \frac{1}{\sqrt{2}}A_{2, 0}^1.
\end{equation}
 In the end, in this tensor-parametrization form of the density matrix,  we are left with two free parameters that can be be determined from data.

\subsection{Quantum entanglement and Bell inequality for two-qutrit system}
To construct the optimized Bell inequality, two observers A and B, each having two measurements—$A_1$ and $A_2$ for $A$, and $B_1$ and $B_2$ for $B$---are considered.
For each value of $A$ and $B$, there are three possible outcomes. One can denote by $P(A_i = B_j+k)$ the probability that the outcomes $A_i$ and $B_j$ differ by $k$ modulo 3. Then we can obtain a simple function by linear sum of  these functions:
\begin{align}
\nonumber
    \mc{I}_3 = &P(A_1 = B_1) + P(B_1 = A_2+1) + P(A_2 = B_2) + P(B_2 = A_1)\\
    &-[P(A_1 = B_1 - 1) + P(B_1 = A_2) + P(A_2 = B_2 - 1) + P(B_2 =A_1 - 1)].
    \label{eq:i3}
\end{align}
$\mc{I}_3$ is bounded as $\mc{I}_3\leq2$ in classical theories as well as other theories complying with local realism~\cite{Collins:2002sun}. 

Once we have determined the coefficients given in Eq.~\ref{eq:ACLM} from the angular distributions of the leptons, one can immediately calculate the density matrix for the composite system, allowing to probe the entanglement and test Bell inequality violation through the expectation values of the Bell operator.
\begin{equation}
    \langle \mc{B}\rangle = \Tr \left[\rho \mc{B}\right],
    \label{eq:bval}
\end{equation}
where $\mc{B}$ is the quantum Bell operator~\cite{Braunstein} for spin-1 massive gauge boson. The construction of this Bell operator  is always based on the linear combination of the three spin-1 component operators or the irreducible tensors~\cite{Fabbri:2023ncz,Barr:2021zcp,Aguilar-Saavedra:2022wam}. It is worth noting that the violation of ~\ref{eq:i3} obtained with the expectation values of the Bell operator may not be the maximal one because in the non-relativistic limit the joint $ZZ$ system is not totally in the singlet state (but close)~\cite{Aguilar-Saavedra:2022wam}. Furthermore, we choose the outgoing direction of the on-shell boson as the polarization axis. However, a unitary rotation on the spin states may maximize the expectation value with following transformation on the Bell operator
\begin{equation}
    \mc{B}\longrightarrow(U\otimes V)^{\dagger}\times\mc{B}\times(U\otimes V),
\end{equation}
where U and V are the independent three-dimensional arbitrary unitary matrices. Here we make use the Bell operator provided in~\cite{Aguilar-Saavedra:2022wam} with the above unitary transformation:
\begin{align}
    \mc{B}=&\left[\frac{2}{3\sqrt{3}}\left( T_{1}^{1}\otimes T_{1}^{1}-T_{0}^{1}\otimes T_{0}^{1}+T_{1}^{1}\otimes T_{-1}^{1} \right) +\frac{1}{12}\left( T_{2}^{2}\otimes T_{2}^{2}+T_{2}^{2}\otimes T_{-2}^{2} \right)\right.\nonumber\\
    &\left. +\frac{1}{2\sqrt{6}}\left(T_2^2\otimes T_0^2+T_0^2\otimes T_2^2\right)-\frac{1}{3}(T_1^2\otimes T_1^2+T_1^2\otimes T_{-1}^2)+\frac{1}{4}T_0^2\otimes T_0^2 \right]+\rm{h.c.}.
\end{align}
Notice that we can use representation of density matrix in Eq.\ref{eq:rho2} and then Bell inequality expectation value is:
\begin{equation}
    \mc{I}_3 = \frac{1}{36}\left( 18+16\sqrt{3}-\sqrt{2}\left( 9-8\sqrt{3} \right) A_{2,0}^{1}-8\left( 3+2\sqrt{3} \right) C_{2,1,2,-1}+6C_{2,2,2,-2} \right) 
\end{equation}

\section{Numerical Simulation and results}

We examine the sensitivity of a future Muon Collider to the entanglement and violation of Bell inequality in $H\to ZZ$ process. The corresponding Higgs boson is produced through VBF process: $W^+ W^-\rightarrow H$. It is worth noting that the signal of this VBF process in a Muon Collider is very clean, almost background free as to be shown below. Here the Higgs boson is on-shell and eventually decays into pair of $Z$ bosons , one of  which  is on-shell while the other is off-shell. We perform a Monte-Carlo simulation of $\mu^+\mu^-\to \nu_{\mu}\bar{\nu}_{\mu} H$, $H \to ZZ^*\to (\ell^+\ell^-)(\ell^+\ell^-)$ events, where $\ell\in\{e, \mu \}$ and $Z^*$ representing the off-shell $Z$ boson, using the \madgraph 
software~\cite{Alwall:2014hca} which includes full spin correlation and Breit-Wigner effects. A sample of one million events are generated at leading order with $\sqrt{s} = 1, 3, 10$ TeV energies for a Muon Collider. These events are then passed to \pythia8.3~\cite{Bierlich:2022pfr} for further event hadronization. Eventually all events are passed through \delphes program~\cite{deFavereau:2013fsa} to include detector effects with a configured file designed for a Muon Collider~\footnote{https://github.com/delphes/delphes/blob/master/cards/delphes\_card\_MuonColliderDet.tcl}. 

The LO cross sections for Higgs production through VBF is relatively large at a TeV Muon Collider~\cite{Forslund:2022xjq}, which is about $1 {\rm pb}$. However, the final cross sections for the considered process is suppressed due to the smaller branching ratio of the Higgs boson into four leptons (muon or electron pairs).  The final leptons  coming from the $Z$ boson pairs can be identified as either four electrons, four muons, or two electrons and two muons. We focus on the $e^+ e^- \mu^+\mu-$ category for simplicity. We choose not to apply cuts on final leptons to avoid possible bias in quantum entanglement measurement when generating parton-level events with \madgraph. On the other hand, cuts on the kinematic variables such as transverse momentum $p_{\rm T}$ and pseudo-rapidity $\eta_{\ell}$ of leptons are applied during the run of \delphes program to mimic the real experiment detector effect. The minimum value of the electron or muon transverse momentum, $p_{\rm T}$, is set to 0.5 GeV, and the maximum value for absolute rapidity, $|\eta_{\ell}|$, is 2.5. 

In the references~\cite{Baranov:2008zzb, Barr:2021zcp,Barr:2024djo}, it is argued that the two $Z$ boson states is not differentiable because of their interaction nature with leptons. However, this difficulty may be avoided in this scenario. The two $Z$ boson can be differentiated. The on-shell $Z$ boson, labeled as $Z_1$, can be identified if the invariant mass of lepton pairs is very close to the true physical mass of $Z$ boson. Similarly, the off-shell $Z$  boson, labeled as $Z_2$, can also be identified if invariant mass of the lepton pair is much smaller than the actual $Z$ mass.label on-shell $Z$ as $Z_1$ whose momentum are determined from summing over its decay products $\ell_1^+ \ell_1^-$, while  the off-shell $Z$ is labeled  $Z_2$ and its momentum are reconstructed from its decay leptons $\ell_2^+\ell_2^-$. We apply identical coordinate system to measure the spin polarization observable as in references~\cite{Bernreuther:2015yna,Aguilar-Saavedra:2022wam}. First of all, a set of coordinates are set up in the Higgs boson rest frame into  which both boson pairs and leptons are boosted.  Finally, the angular coordinates of the leptons, $(\theta_1^-, \phi_1^-)$ for negative-charged lepton from $Z_1$ decay and $(\theta_2^-, \phi_2^-)$ for negative-charged lepton from $Z_2$ decay can be obtained and used to determine the coefficients for density matrix in Eq.~\ref{eq:rho2}. through the integration given in~\ref{eq:ACLM}.

In order to adjust our simulation to the real experiment, we customize the whole events to fit to the number of achievable events in a muon collider based on the integrated luminosity. Tab.~\ref{tab:x_section} shows the cross sections given by the \madgraph event generator and the events corresponding to the luminosity and cross sections.

\begin{table}[h]
\captionsetup{justification=raggedright,singlelinecheck=false}
    \centering
    \begin{tabular*}{0.6\textwidth}{@{\extracolsep{\fill}} cccc}
        \toprule
        $\sqrt{s_{\mu}}[{\rm TeV}]$ & $\sigma[{\rm fb}]$         &    Luminosity  &  Events \\ 
         \hline\hline
         1  & $1.51\times 10^{-2}$ &    30 ${\rm ab}^{-1}$  &  455  \\ \hline
         3  & $3.56\times 10^{-2}$ &    30 ${\rm ab}^{-1}$  &  1089  \\ \hline
         10 & $6.06\times 10^{-2}$ &    30 ${\rm ab}^{-1}$  &  1890  \\ 
        \bottomrule
    \end{tabular*}
  \caption{The total cross sections given by the generator at the three different CM energies and the number of events achievable by a muon collider with integrated luminosity ${\rm L} = 30~{\rm ab^{-1}}$}.
    \label{tab:x_section}
\end{table}

\begin{figure}
    \centering
    \includegraphics[width=0.45\linewidth]{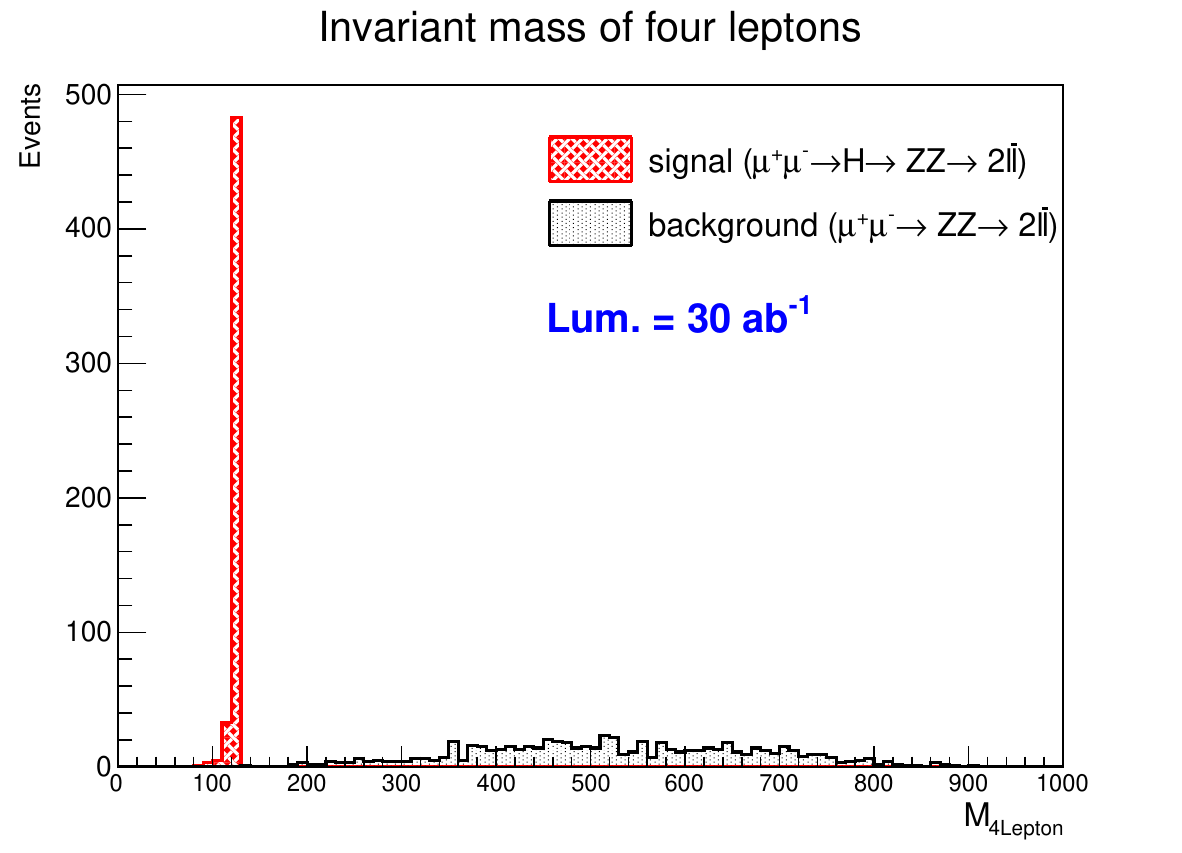}
    \caption{Signal and background distribution events as the four lepton invariant mass. Figure on the left panel shows $M_{4\ell}$ for total one million events while right panel shows that invariant mass distribution corresponding to the assumed luminosity of 30 ${\rm ab}^{-1}$.}
    \label{fig:bkg}
\end{figure}

\begin{figure}[h!]
    \centering
    \includegraphics[width=0.45\linewidth]{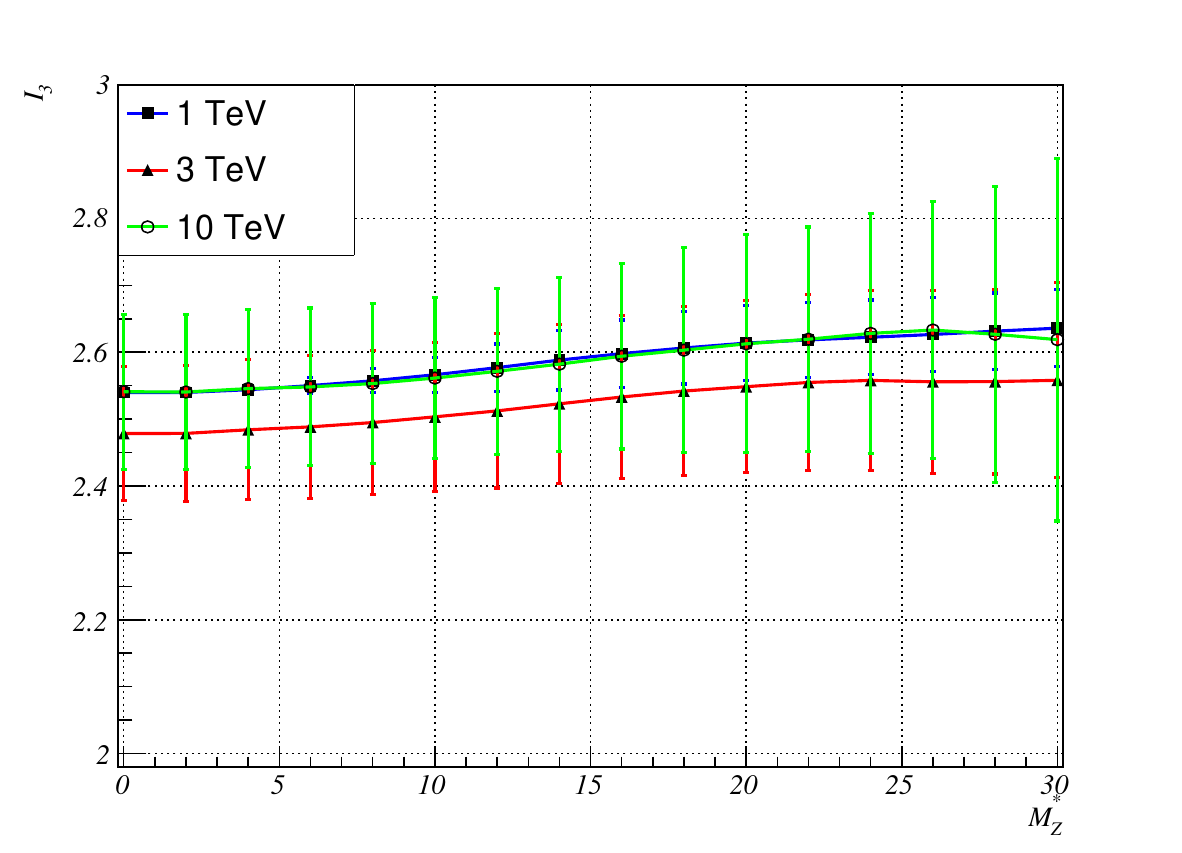}
    \caption{The expectation value of Bell inequality as function of the off-shell Z mass $M_{Z_2}$. $I_3$ is measured at three CM energy. Points represent mean value of $I_3$.}
    \label{fig:fig2}
\end{figure}
Backgrounds at muon colliders are generally several orders of magnitude smaller than at hadron colliders~\cite{AlAli:2021let}.
We expect the electroweak process $\mu^+\mu^-\to ZZ\to 4\ell$ to be the main background in our analysis, where the two Z bosons are produced through W boson fusion. The four-lepton invariant mass distributions for the signal and background are shown in Fig.~\ref{fig:bkg}. The red-filled area depicts our signal events while the black-filled area represents the background events. The signal events are clearly distinguishable from the signal events, making it a very good approximation to assume that there is no significant background for our Higgs signal.

The statistical uncertainties of the  coefficients $\cc,~\CC$ and $I_3$ are estimated through slicing the total one million events into number of pseudo-experiments, and calculate the observables from the differential distribution. Each pseudo-experiment includes the number of events corresponding to the achievable events for target luminosity and cross section shown in Tab.~\ref{tab:x_section}. So that we will have approximately 500 $\sim$ 2000 pseudo-experiments. Repeating the calculation over these pseudo-experiments will give us the mean and standard deviation for each observable. Here, the standard deviation is understand as statistical uncertainty. Because of the clean final states and fine experimental resolutions for light charged leptons, systematic uncertainty are not considered in this study. As stated in previous sections, the entanglement strength is relevant with the off-shell Z boson mass: the larger the mass $M_{Z_2}$, the more entangled the $Z$ boson pairs are, and the greater the value of $I_3$. This feature is shown in the figure~\ref{fig:fig2}. Here the data points represent the mean value of the Bell operator at the specific off-shell Z mass with corresponding CM energies. The blue line depicts the $I_3$ value extracted from data with $\sqrt{s} = 1$ TeV, while the red and green lines represent $I_3$ determined from data with respect 3 and 10 TeV CM energies. Clearly, the Bell inequality is  significantly violated, in contrast to to the predictions of the classical deterministic theories. The actual lower limit of the off-shell Z mass can be larger than 30 GeV. However, putting a to much  mass limit on $M_{Z_2}$ would increase the statistical uncertainty in the measurement, producing low significance. So we are limited with 30 GeV mass-limit for the off-shell boson mass.
\begin{table}[h!]
\centering
$\sqrt{s} = 1$ TeV\\
\begin{tabular*}{0.6\textwidth}{@{\extracolsep{\fill}} cccc}
\toprule
$M_{Z_2}$ (GeV) & $I_3$ & $\cc$ & $\CC$ \\
\midrule
0.000 & $2.563 \pm 0.325$ & $-0.928 \pm 0.216$ & $0.527 \pm 0.164$ \\
\hline
10.000 & $2.596 \pm 0.335$ & $-0.943 \pm 0.220$ & $0.553 \pm 0.179$ \\
\hline
20.000 & $2.654 \pm 0.373$ & $-0.977 \pm 0.248$ & $0.574 \pm 0.192$ \\
\hline
30.000 & $2.663 \pm 0.508$ & $-0.979 \pm 0.334$ & $0.589 \pm 0.248$ \\
\bottomrule
\end{tabular*}
\caption{Values of the correlation coefficients $\cc$ and $\CC$ as the signal for quantum entanglement and also the expectation value of the Bell operator $I_3$. The expected target luminosity is $30 {\rm ab}^{-1}$ and $\sqrt{s} = 1$ TeV.}
\label{tab:I3_1tev}
\end{table}

\begin{table}[h!]
\centering
$\sqrt{s} = 3$ TeV\\
\begin{tabular*}{0.6\textwidth}{@{\extracolsep{\fill}} cccc}
\toprule
$M_{Z_2}$ (GeV) & $I_3$ & $\cc$ & $\CC$ \\
\midrule
0.000 & $2.467 \pm 0.217$ & $-0.871 \pm 0.121$ & $0.493 \pm 0.377$ \\
\hline
10.000 & $2.499 \pm 0.225$ & $-0.891 \pm 0.135$ & $0.502 \pm 0.390$ \\
\hline
20.000 & $2.538 \pm 0.254$ & $-0.908 \pm 0.163$ & $0.536 \pm 0.365$ \\
\hline
30.000 & $2.543 \pm 0.342$ & $-0.890 \pm 0.216$ & $0.606 \pm 0.423$ \\
\bottomrule
\end{tabular*}
\caption{Same as \ref{tab:I3_1tev} but for $\sqrt{s} = 3$ TeV.}
\label{tab:I3_3tev}
\end{table}

\begin{table}[h!]
\centering
$\sqrt{s}=10$ TeV\\
\begin{tabular*}{0.6\textwidth}{@{\extracolsep{\fill}} cccc}
    \toprule
    $M_{Z_2}$ (GeV) & $I_3$ & $\cc$ & $\CC$ \\
    \midrule
    0.000 & 2.539 $\pm$ 0.312 & -0.930 $\pm$ 0.196 & 0.466 $\pm$ 0.232 \\
    10.000 & 2.569 $\pm$ 0.295 & -0.946 $\pm$ 0.194 & 0.482 $\pm$ 0.217 \\
    20.000 & 2.616 $\pm$ 0.321 & -0.969 $\pm$ 0.218 & 0.514 $\pm$ 0.219 \\
    30.000 & 2.644 $\pm$ 0.517 & -0.943 $\pm$ 0.334 & 0.527 $\pm$ 0.280 \\
    \bottomrule

    \end{tabular*}
\caption{Same as \ref{tab:I3_1tev} and \ref{tab:I3_3tev} but for $\sqrt{s} = 10$ TeV.}
\label{tab:I3_10tev}
\end{table}
The final measurements of the observable quantities are shown in ~\ref{tab:I3_1tev}, \ref{tab:I3_3tev} and \ref{tab:I3_10tev} with corresponding CM energies at the four different  lower mass limits $M_{Z_2}\in[0, 10, 20, 30]$ GeV. The value of $I_3$ becomes larger as expected because larger $M_{Z_2}$ means the joint states entangled more. However, the statistical uncertainties rise as the  $M_{Z_2}$ mass gets larger. Either non-zero value of the correlation coefficients indicates that the two Z boson states are entangled. Quantum entanglement can be probed up to $4\sigma$ of significance with lower $M_{Z_2}$ cut or $2\sigma\sim3\sigma$ with higher $M_{Z_2}$ cut, using either one of  the correlation coefficients $\cc$ and $\CC$. The significance of the violation of Bell inequality can be obtained up to $2\sigma$.

\section{Summary and conclusion}

In this work, we have explored the prospects for probing quantum entanglement and violation of the optimized Bell inequality, CGLMP inequality, in $H\to ZZ\to4 \ell$ process at a future Muon Collider. When the Higgs boson is produced via vector boson fusion, the resulting final state is essentially background-free. Simulations are performed at three center-of-mass energies, $\sqrt{s}$ = 1, 3, and 10 TeV, with assumed integrated luminosity of $30\rm{ab^{-1}}$. Because, the Higgs boson is a spin-zero particle, the $ZZ$ system form a spin-singlet state, significantly reducing the number of free parameters in the spin-density matrix. By measuring spin-correlation coefficients, we are able to not only test for the presence of quantum entanglement but also prove a much more stringent condition: the violation of  Bell inequality.

The signal events are selected by focusing on two electron and two muon channels coming from $ZZ$ decay.  Both the on-shell and off-shell $Z$ bosons are identified by determining the invariant mass of these lepton pairs.  The Z boson with largest or close to the real mass is tagged as on-shell, while the other one, with very small mass, is tagged as off-shell. This method is very unique compared to Z pair production in other processes or colliders. We also included  the detector effects in our simulation using \delphes program configured for a muon detector.  Cuts are being applied on the lepton kinematic variables $p_{\rm T}$ and $|\eta|$ based on the current muon detection technology.  
We have also simulated ZZ pair production through $W$ fusion as background to our signal and found out that it to be negligible.

The density matrix of the ZZ state is expressed with a simple yet very effective parametrization~\cite{Aguilar-Saavedra:2022wam} which contains only two independent parameters. In the end, we are left with spin-correlation coefficients $\cc$ and $\CC$ that can determine the density matrix.  These coefficients can be extracted from actual or simulated data by determining the spherical coordinates of the final leptons, namely electron and muon. Any non-zero value of $\cc$ or $\CC$ signals quantum entanglement between paired Z boson states. Finally, we found that entanglement can be probed with a significance of around $4\sigma$ and the violation of the Bell inequality can be tested up to $2\sigma$ level.  

\begin{acknowledgments}
This work is supported in part by the National Natural Science Foundation of China under Grants No. 12325504, No. 12150005, and No. 12075004. 
\end{acknowledgments}

\section*{Appendix}

\bibliographystyle{ieeetr}
\bibliography{paper_ref}

\end{document}